\documentclass[prl,aps,floatfix,tightenlines,twocolumn,superscriptaddress,amsmath,amssymb,letterpaper]{revtex4-1}

\usepackage{graphicx}
\usepackage{dcolumn}
\usepackage{bm}
\usepackage{epsfig}
\def\comment#1{}

\begin{document}
\title{Coherent coupling of a superconducting flux-qubit to an electron spin ensemble in diamond}

\author{Xiaobo Zhu}
\affiliation{NTT Basic Research Laboratories, NTT Corporation, 3-1 Morinosato-Wakamiya, Atsugi, Kanagawa 243-0198, Japan}

\author{Shiro Saito}
\affiliation{NTT Basic Research Laboratories, NTT Corporation, 3-1 Morinosato-Wakamiya, Atsugi, Kanagawa 243-0198, Japan}

\author{Alexander Kemp}
\affiliation{NTT Basic Research Laboratories, NTT Corporation, 3-1 Morinosato-Wakamiya, Atsugi, Kanagawa 243-0198, Japan}

\author{Kosuke Kakuyanagi}
\affiliation{NTT Basic Research Laboratories, NTT Corporation, 3-1 Morinosato-Wakamiya, Atsugi, Kanagawa 243-0198, Japan}

\author{Shin-ichi Karimoto}
\affiliation{NTT Basic Research Laboratories, NTT Corporation, 3-1 Morinosato-Wakamiya, Atsugi, Kanagawa 243-0198, Japan}

\author{Hayato Nakano}
\affiliation{NTT Basic Research Laboratories, NTT Corporation, 3-1 Morinosato-Wakamiya, Atsugi, Kanagawa 243-0198, Japan}

\author{William J. Munro}
\affiliation{NTT Basic Research Laboratories, NTT Corporation, 3-1 Morinosato-Wakamiya, Atsugi, Kanagawa 243-0198, Japan}

\author{Yasuhiro Tokura}
\affiliation{NTT Basic Research Laboratories, NTT Corporation, 3-1 Morinosato-Wakamiya, Atsugi, Kanagawa 243-0198, Japan}

\author{Mark S. Everitt}
\affiliation{National Institute of Informatics, 2-1-2 Hitotsubashi, Chiyoda-ku, Tokyo-to 101-8430, Japan}

\author{Kae Nemoto}
\affiliation{National Institute of Informatics, 2-1-2 Hitotsubashi, Chiyoda-ku, Tokyo-to 101-8430, Japan}

\author{Makoto Kasu}
\affiliation{NTT Basic Research Laboratories, NTT Corporation, 3-1 Morinosato-Wakamiya, Atsugi, Kanagawa 243-0198, Japan}

\author{Norikazu Mizuochi}
\affiliation{University of Osaka, Graduate school of Engineering Science, 1-3, Machikane-yama, Toyonaka, Osaka, 560-8531, Japan}
\affiliation{PRESTO JST, 4-1-8 Honcho, Kawaguchi, Saitama 332-0012, Japan}

\author{Kouichi Semba}\email{semba.kouichi@lab.ntt.co.jp}
\affiliation{NTT Basic Research Laboratories, NTT Corporation, 3-1 Morinosato-Wakamiya, Atsugi, Kanagawa 243-0198, Japan}

\begin{abstract}

Electron-spin nitrogen-vacancy color centers in diamond are a natural  candidate to act as a  quantum memory for superconducting qubits because of their large collective coupling and long coherence times. We report here the first demonstration of  strong coupling and coherent exchange of a single quantum of energy between a flux-qubit and an ensemble of nitrogen-vacancy color centers.  
\end{abstract}

\date{\today}

\maketitle

During the last decade, research into superconducting quantum bits (qubits) based on 
Josephson junctions has made rapid progress \cite{CW08}. Many 
foundational experiments have been performed 
\cite{Nakamura99,Vion02,Irinel03,NIST07,Yale07,DiCarlo09,Ansmann09,Neeley10,DiCarlo10} 
and superconducting qubits are now considered 
one of the most promising systems for quantum information processing.  However,  the 
experimentally reported coherence times are likely to be insufficient for future 
 large scale quantum computation. A natural solution is a dedicated {\it engineered} 
 quantum memory based on atomic and molecular systems. Since macroscopic quantum coherence 
 was first demonstrated in Josephson junction circuits \cite{Nakamura99}, the 
 question of whether or not coherent quantum coupling between a single macroscopic artificial 
atom and an ensemble of natural atoms or molecules is possible 
has attracted significant attention \cite{Sorensen04,Tian04,Rabl06,Marcos10}.
In this letter we present for the first time evidence of coherent strong coupling between a single macroscopic superconducting artificial atom 
 (a flux qubit) and an ensemble of electron-spin nitrogen-vacancy color centers (NV$^{-}$ centers) 
 in diamond. Furthermore, we have observed coherent exchange of a single quantum of energy 
 between a flux qubit and  a macroscopic ensemble consisting of $\sim 3\times10^{7}$  of NV$^{-}$ centers. 
 This provides a foundation for future quantum memories and hybrid devices coupling 
microwave and optical systems.


With the early successes of single atom quantum state manipulation \cite{Brune96}, 
research in quantum information processing with atomic and solid-state systems  has progressed 
largely in a separate fashion. In recent years,  significant effort has been devoted to coupling atomic and 
molecular system to solid-state qubits to form hybrid quantum devices \cite{Sorensen04,Tian04,Rabl06}.  
Hybrid devices involving the integration of an atomic system with a superconducting transmission line 
resonator have been realized \cite{Schuster10,Kubo10,TU-Wien11}. Such schemes have the 
potential to couple superconducting solid-state qubits to optical fields via atomic systems, thus enabling
quantum media conversion. The coupling strength $g$ of an individual atomic system to one electromagnetic mode in a resonator 
circuit is usually too small for the coherent exchange of quantum information. 
However, the coupling strength of an ensemble of $N$ such atomic systems will be enhanced by a factor of $\sqrt{N}$ \cite{Kimble89}, 
allowing one to reach the strong coupling regime ($g\sqrt{N}\gg \kappa, \gamma$, 
where $\kappa$ and $\gamma$ are the damping rates of resonator circuit and atomic system). 

Of the many possible hybrid systems,  coupling a flux-qubit to 
an (NV$^{-}$) center in diamond is particularly appealing. 
Firstly, the magnetic coupling between a flux-qubit and a single NV$^{-}$ center 
can be three orders of magnitude larger than that associated 
with a superconducting transmission line resonator \cite{Marcos10}. 
Second, the ground state of an NV$^{-}$ center is a spin 1 triplet due to its  
 $C_{3v}$ symmetry  (Figure 1b). The $S=1$ spin triplet $|m_{s}=0\rangle$ 
state is separated by 2.88 GHz from the near degenerate excited states 
$|m_{s}=\pm 1\rangle$ under zero magnetic field (Figure 1c). This energy 
separation is ideal for a gap tunable flux-qubit to be brought on and off resonance with it.

\begin{figure}[htb]
\begin{center}
\includegraphics[width=1.0\columnwidth,clip]{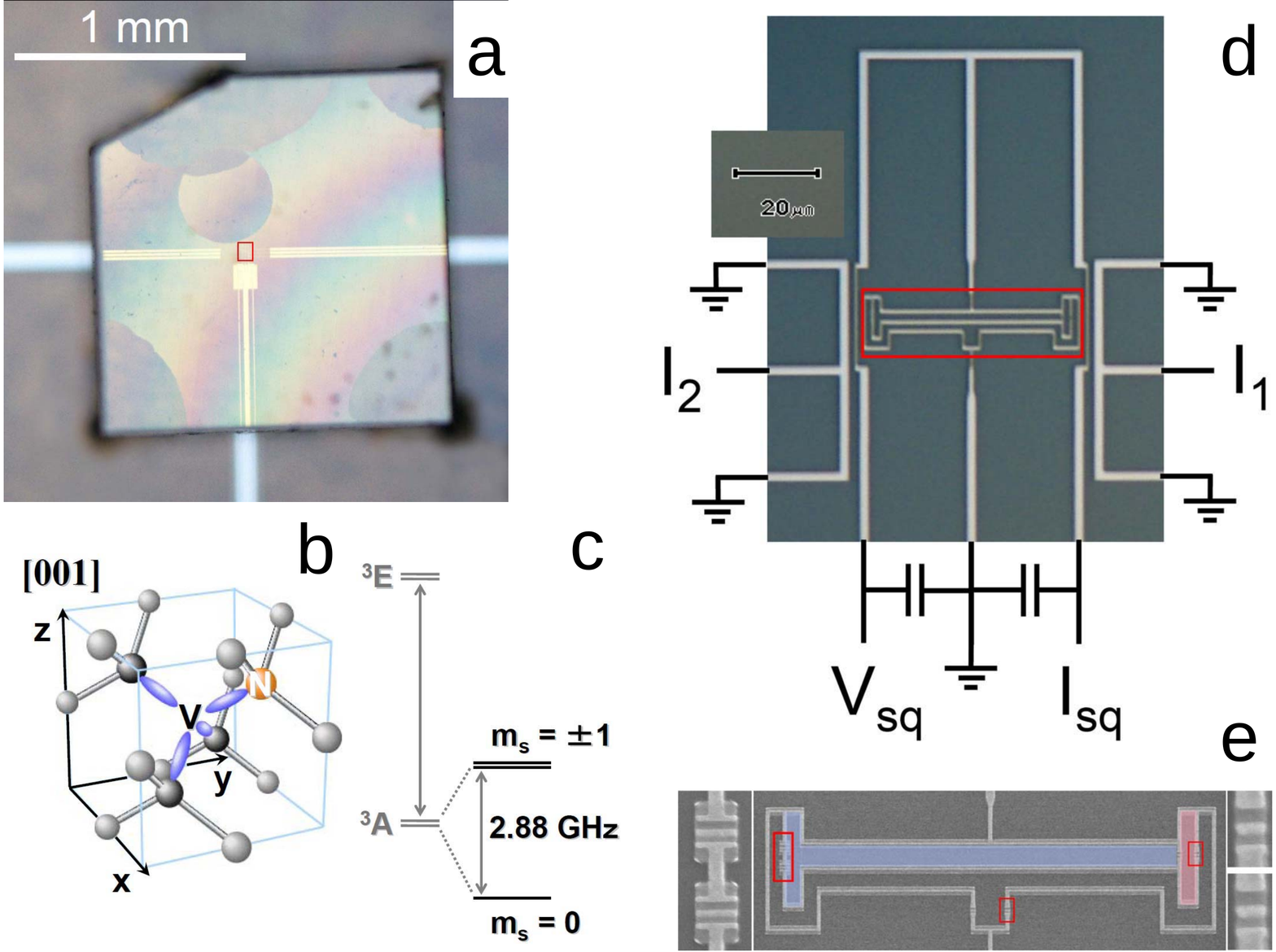}
\caption{
{\bf Experimental set-up of an NV-diamond sample attached to a flux-qubit system. }
(a) A diamond crystal is glued on top of a flux-qubit and its superconducting circuits
 (under the red box) with the diamonds $^{12}$C implanted (001) surface facing the chip. 
The distance between the flux-qubit and surface of the diamond crystal is 
carefully adjusted to be less than a micrometer using 100 nm height mesa 
structure on the diamond surface, a circle adjacent to the red square, 
and the optical interference pattern (Newton's ring).
(b) A sketch of a NV color center of diamond with its vacancy (V) and 
nitrogen (N) atom, as well neighboring carbon atoms. Four equivalent 
NV-axis exist depicted in purple color dangling bonds, all making 
the same angle with [001] direction to which the magnetic field 
generated by the flux-qubit points. 
(c) Energy diagram of the NV center, with the spin triplet $|m_{s}=0\rangle$ 
ground state separated by 2.88 GHz from the degenerated 
$|m_{s}=\pm 1\rangle$ excited states under zero magnetic field\cite{Jelezko04}. 
(d) Optical micrograph and the circuit scheme of the aluminum made 
flux-qubit, the magnified view of the chip under the red box region 
shown in Fig1a. The central M shaped circuit contains a flux-qubit and
 a SQUID detector. Two high-bandwidth (20 GHz) MW-control lines
 located both sides of the qubit circuit. 
(e) The H-shaped gap tunable flux-qubit and the edge-shared SQUID 
used as a switching qubit state detector. The flux-qubit contains two loops, 
the main loop (blue) and the $\alpha$-control loop (magenta) 
which controls tunneling energy gap of the flux-qubit. 
Magnified view of Josephson junctions are also shown. 
The mutual inductance of control line-1 to the $\alpha$-loop and 
main loop are 90 fH and 256 fH, and those of control line-2 
are 0.5 fH and 549 fH. The magnetic flux penetrating these 
two loops can be controlled in ns time scale by applying synchronized 
current pulses to these control lines in situ.}
\label{FIG_1}
\end{center}
\end{figure}

In this Letter, we report on the first observation of vacuum Rabi oscillations 
between a flux-qubit and an ensemble of approximately 
three million NV$^{-}$ centers in diamond. This demonstrates strong 
coherent coupling between two dissimilar quantum systems 
with an effective collective coupling constant of $g_{\rm ens}\sim70$ MHz.

We begin by describing our experimental setup  as depicted  in (Figure 1).  
An NV$^{-}$ diamond sample was prepared  by ion implantation of $^{12}C^{2+}$ at 
700 keV under high vacuum into high-pressure, high-temperature (HPHT)-synthesized type Ib (001) surface orientation single
crystal diamond. The $^{12}C^{2+}$ ions, with a dose condition of $3\times10^{13}$ cm$^{-2}$, 
were stopped at a depth of $ 600^{+50}_{-100}$ nm. This generated on the order of $5\times10^{18}$ cm$^{-3}$ 
vacancies over a depth of $\sim0.7 \mu$m. 
After implantation, the crystals were annealed at $900 ^{\circ}$C under vacuum for 3 hours. 
This high dose carbon implantation method enhances the yield of generated NV$^{-}$ centers  \cite{Naydenov10}.
Photoluminescence (PL) optical spectroscopy (shown in Figure 2) established that NV$^{-}$ centers 
were generated with a density of $\sim 1.1\times10^{18}$ cm$^{-3}$  over a $1 \mu$m 
depth.  We can describe the ground state of a single NV$^{-}$ center by the Hamiltonian \cite{Neumann09},
\begin{eqnarray}
H_{ \rm  NV}=hDS_{z}^2+hE(S_{x}^2-S_{y}^2)+ h g_{\rm NV}\mu_{\rm B}{\bf B}\cdot{\bf S},
\end{eqnarray}
where $S_x, S_y, S_z$ are the usual Pauli spin 1 operators,  $D$ the zero-field splitting  (2.878 GHz), 
$E$ the strain-induced splitting ($<$1 MHz), the N-V Land\'{e} factor $g_{NV}=2$, 
and $\mu_{B}=14$ MHz$/$mT.  The last term represents the Zeemen splitting, which is negligible 
in our case as the magnetic field applied perpendicular to the surface of the chip to prepare the flux qubit is 
less than $0.1$ mT.

\begin{figure}[htb]
\begin{center}
\includegraphics[width=0.87\columnwidth,clip]{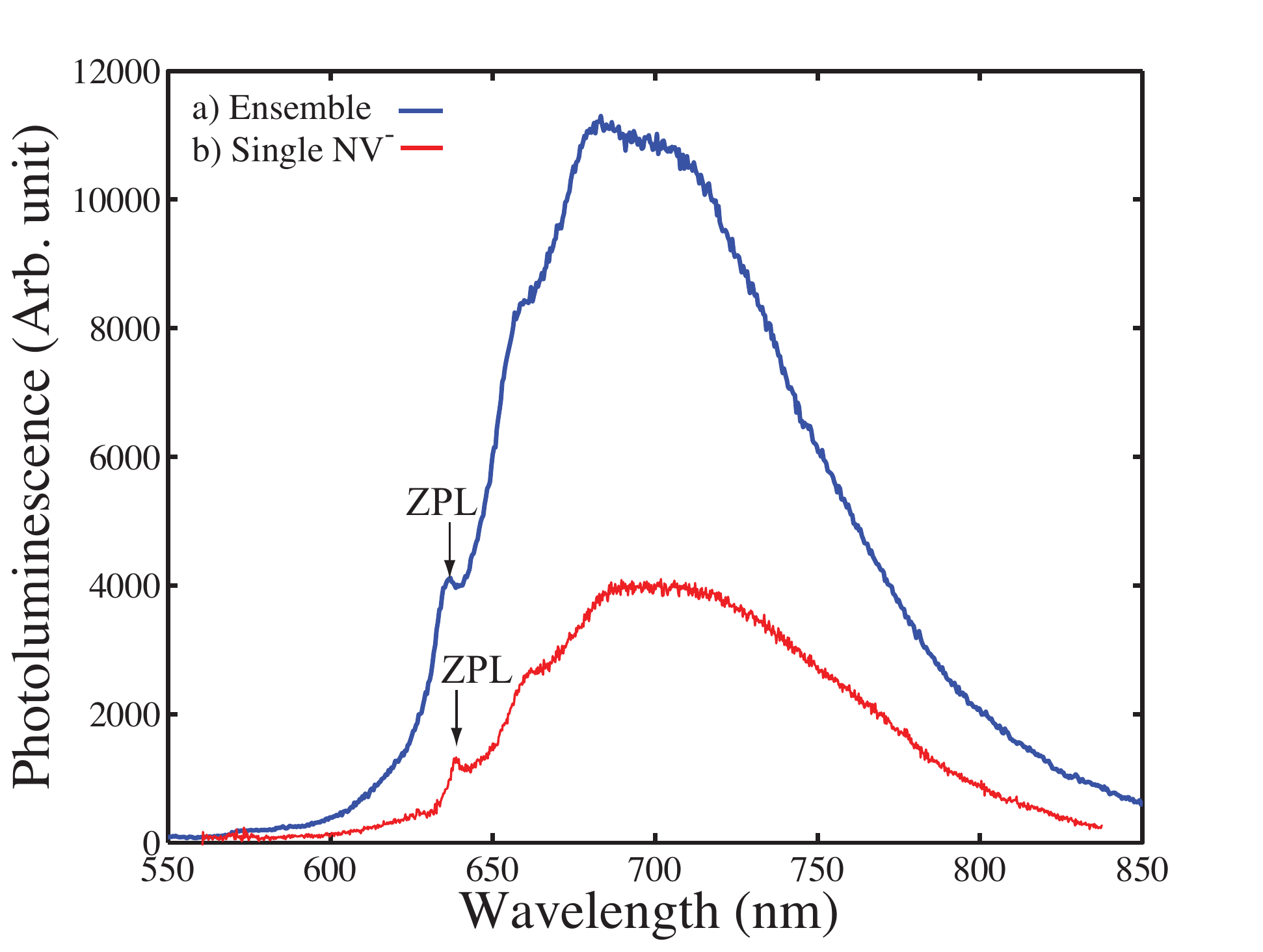}
\caption{
{\bf Photoluminescence spectra } of (a) ensembles of color centers in the 
highly carbon implanted sample and (b) a single NV- center in pure diamond 
at room temperature. Their signal intensities are normalized for the comparison 
of the spectra. In the spectrum of the single NV center, the contributions of 
phonon Raman scattering from bulk diamond at 573 and between 600 and 620 nm 
are subtracted\cite{Gru}. The zero phonon line of NV- at 637 nm\cite{Gru,Kur} 
is clearly observed in both spectra. Broad spectrum of phonon replicas is also 
very similar to each other and to the reported ones\cite{Gru,Kur}.These indicate 
that the NV- center is produced as a major color center in the highly carbon 
implanted sample. The signal intensity of the ensemble is about $6.5\times10^{4}$ 
times stronger than that of the single one. From this result, the concentration 
of the NV- center in the highly carbon implanted sample was estimated to be 
$1.1\times10^{18}$ cm$^{-3}$. }
\label{FIG_2}
\end{center}
\end{figure}

A diamond crystal was glued on top of the superconducting circuit
with the $^{12}$C implanted surface facing the flux-qubit (Figure 1a). 
We used a gap tunable flux-qubit \cite{Zhu10,Fedorov10} (Figure 1d, 1e) where the smallest junction 
of the three Josephson junction qubit is replaced by a low inductance 
superconducting quantum interference device (dc-SQUID) loop (the magenta loop in Figure 1e).
The flux qubit - NV ensemble coupled system is measured by the qubit state using a 
built-in dc-SQUID (the biggest loop in Fig.1e sharing edges with the flux-qubit) 
which is inductively coupled to the qubit. 
When biasing the main loop close to half a flux quantum, the device is 
an effective two-level system \cite{Mooij99} described by the Hamiltonian
\begin{eqnarray} 
H_{\rm qb}=\frac{h}{2}(\Delta\sigma_{x}+\epsilon\sigma_{z}),
\end{eqnarray}
which is given in the basis of clockwise and counter-clockwise currents. Here, $\sigma_{x,z}$ are the Pauli 
spin $\frac{1}{2}$ matrices, $h\epsilon=2I_{\rm P}(\Phi_{\rm ex}-3\Phi_{0}/2)$ is the energy 
bias ($I_{\rm P}\approx300$ nA is the persistent current in the qubit, $\Phi_{\rm ex}$ is the external flux 
threading the qubit loop, and $\Phi_{0}=h/2e$ is the flux quantum), and $\Delta$ 
is the tunnel splitting. The energy splitting of the gap tunable flux-qubit is 
$hF=h\sqrt{\epsilon^{2}+\Delta^{2}}$ where $\epsilon$ and $\Delta$ can be 
controlled independently by the external magnetic flux threading the two loops.
This type of flux-qubit can be tuned into resonance with an NV$^{-}$ 
ensemble in-situ at a base temperature of  $\sim$12 mK while 
keeping the qubit at its optimum flux bias (degeneracy point). 
The total Hamiltonian of the coupled system is
\begin{eqnarray} 
H&=&
\frac{h}{2}(\Delta\sigma_{x}+\epsilon\sigma_{z}) \nonumber \\
&+& h \sum_{i} \left[ D S_{z,i}^2 + E (S_{x,i}^2-S_{y,i}^2)\right]\nonumber \\
&+& \frac{h}{2} \sum_i g_i \sigma_z S_{x,i}, 
\end{eqnarray} 
where $i$ runs over the NV$^{-}$ centers which couple to the flux qubit. The corresponding 
coupling  constant can be estimated using the Biot-Savart law at $g_i \sim 8.8$ kHz. In our situation 
here, the $|\pm 1\rangle_i$ states of  the NV$^{-}$ electronic spin are near degenerate and so our flux-qubit 
couples to both the $\left|0\right\rangle_i \rightleftharpoons\left|1\right\rangle_i $ and 
$\left|0\right\rangle_i \rightleftharpoons\left|-1\right\rangle_i$ transitions. This results in an effective coupling constant 
$\sqrt{2} g_i$ larger that generally anticipated. 

\begin{figure}[htb]
\begin{center}
\includegraphics[width=0.95\columnwidth,clip]{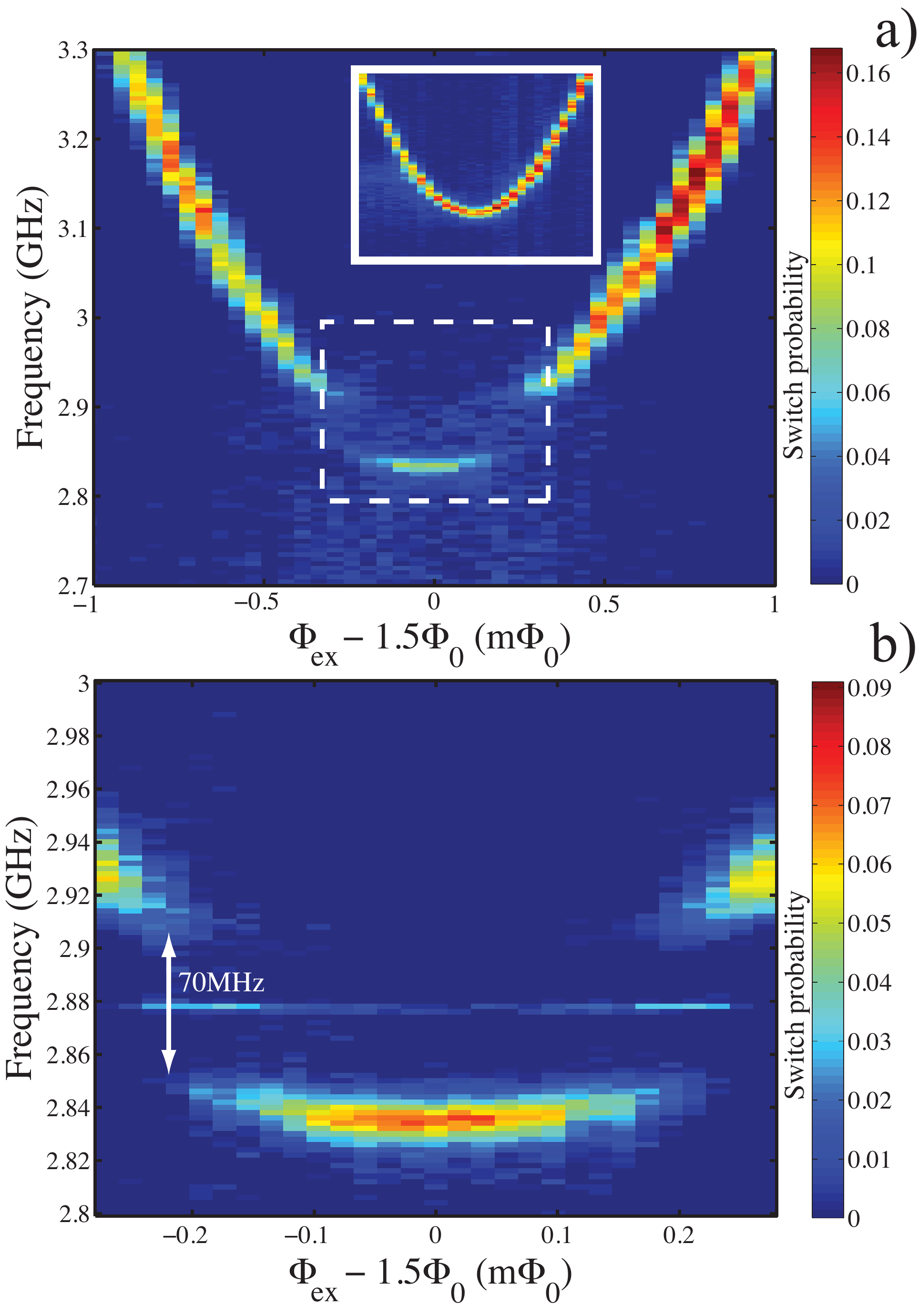}
\caption{
{\bf Energy spectrum of the flux-qubit coupled to ensemble of NV-centers.}
a) Resonant frequencies indicated by peaks in the SQUID detector switching probability 
when a 500 ns long microwave pulse excites the system before the readout pulse. 
Data are represented as a function of the external magnetic flux through the effective 
qubit area ($\Phi_{ex}=\Phi_{m}+\Phi_{\alpha}/2$, where $\Phi_{m}$ and $\Phi_{\alpha}$ 
are the flux through the qubit main-loop and $\alpha$-loop, respectively). 
Inset:  resonant frequency spectrum of the same flux-qubit over the same region without mounting the NV-diamond crystal. 
(b) Magnified view of the white dotted box region in (a). A vacuum Rabi splitting as large as $70$ MHz is clearly observed. 
Since the qubit phase relaxation time measured by spin-echo $T_{2}^{\rm echo}\approx0.25$ ${\mu}s$, 
the strong coupling condition is satisfied by the $\sqrt{N}$ enhancement in $g$. $N\approx 3\times10^{7}$ 
is the estimated number of electron spins in the ensemble strongly coupled to the flux-qubit.}
\label{FIG_3}
\end{center}
\end{figure}

\begin{figure}[!htb]
\begin{center}
\includegraphics[width=0.95\columnwidth,clip]{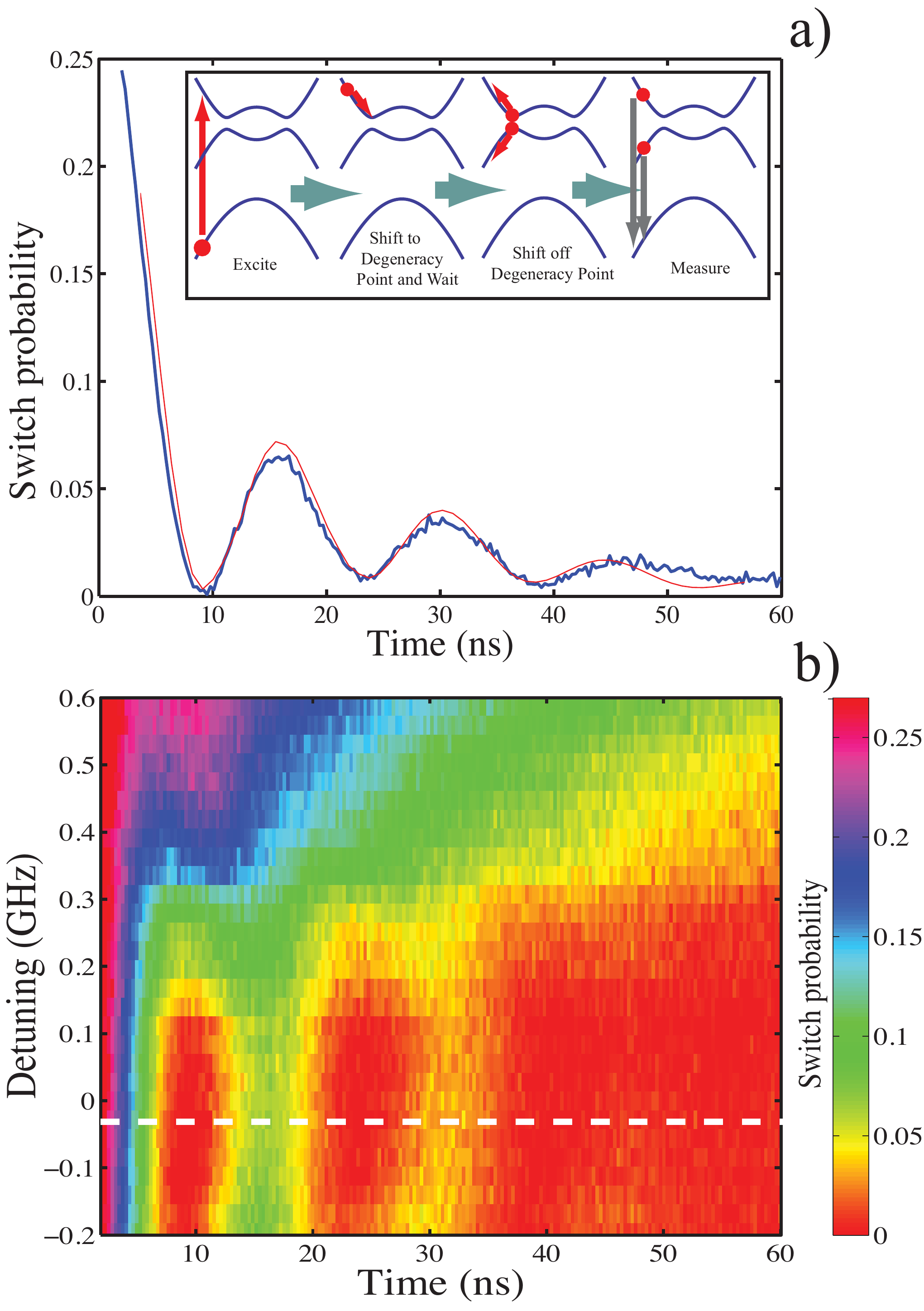}
\caption{{\bf Vacuum Rabi oscillation of the flux-qubit NV-ensemble coupled system.}
(a) Damped oscillation (blue curve) of an initially excited flux coupled to an NV-ensemble with 
the results from a phenomenological model shown as a thin red curve. 
Inset, schematic measurement sequence for the vacuum Rabi measurement. An Initial ground 
state $|0\rangle_{\rm qubit}|0\rangle_{\rm ens}$ is prepared by letting the system to relax longer than 
800 $\mu$s under the base temperature (T=12 mK) of the dilution refrigerator at the optimal 
readout flux bias. Immediately following this a $\pi$-pulse is applied to the flux qubit resulting in 
the state   $|1\rangle_{\rm qubit}|0\rangle_{\rm ens}$. The system is then brought into resonance 
non-adiabatically by a flux-bias shift current pulse through the control line 2. On resonance 
vacuum Rabi oscillation takes place for a given time;  $|1\rangle_{\rm qubit}|0\rangle_{\rm ens} \rightleftharpoons 
|0\rangle_{\rm qubit}|1\rangle_{\rm ens}$. Finally, the SQUID detector reads the qubit state as a 
function of the time keeping the system on resonance. We repeat these measurement typically 
$2\times10^4$ times under the same condition to obtain good statistics.
(b) 2D plot of the SQUID detector switching probability as a function of both the detuning by flux bias shift 
and the time keeping the flux-qubit NV-ensemble system at a given detuning.  
The white broken line corresponds to the switching probability shown in a). }
\label{FIG_4}
\end{center}
\end{figure}

From the spectroscopic measurements, a clear anti-crossing was observed (Figure 3a) 
near the degeneracy point of the flux-qubit, while no gap was observed 
in the same flux-qubit prior to the mounting of the ensemble (inset in Figure 3a).
We also note  a narrow resonance at 2.878\,GHz of less than 1\,MHz width near these anti-crossings.
This can be ascribed to the near degenerate excited states of the NV$^{-}$ ensemble and so 
indicates a strain-induced zero-field splitting coefficient $E$ of less than 1\,MHz. 
From the fine scan spectroscopy shown in (Figure 3b), a vacuum Rabi splitting near $g_{\rm ens} \sim 70$\,MHz was 
clearly observed confirming strong coupling between the flux-qubit and the NV$^{-}$ ensemble. 
Next from the measured vacuum Rabi splitting and our calculated value of $g_i$ we can estimate 
the number of NV$^{-}$ centers in the ensemble at $N=g^{2}_{\rm ens}/2g^{2}\approx 3.2\times10^{7}$, 
where the factor of $2$ in the denominator is due to the two-fold degeneracy of the excited $|\pm 1\rangle_i$ 
states of an NV$^{-}$ center. This estimate is consistent with the density of NV$^{-}$ centers measured by PL spectroscopy in the whole 
sample ($1.1\times10^{18}$\,cm$^{-3}$) multiplied by the volume of centers coupling to the flux qubit 
(area 40\,$\mu$m$^{2}\times$ effective thickness 0.7\,$\mu m$). The PL spectroscopy approach 
gives the number of coupled centers as $\approx3.1\times10^{7}$.

Next, we investigated the dynamics of our system in the time domain
using a similar measurement cycle to that performed in qubit-LC resonator coupled systems \cite{Jan06}.
We first excited the flux-qubit and then brought it into resonance with the NV$^{-}$ ensemble.
Single energy quantum exchange between the flux-qubit and NV$^{-}$ ensemble at resonance manifests 
itself as the vacuum Rabi oscillations
\begin{equation}
|1\rangle_{\rm qb}|0\rangle_{\rm ens} \rightleftharpoons |0\rangle_{\rm qb}|1\rangle_{\rm ens}
\end{equation}
where $|1\rangle_{\rm ens}=\frac{1}{\sqrt{N}}\sum_{i}  S_{+,i}|00\cdots0\rangle$, with 
$S_{+,i}=|1\rangle_i \langle 0|_i+|-1\rangle_i \langle 0 |_i$ being the raising operator of the 
$i$-th NV$^{-}$ spin to both the $|\pm 1\rangle$ states.  (Figure 4a) clearly shows vacuum Rabi 
oscillations between the flux-qubit and ensemble of electronic spins at the 2.878 GHz resonance. 
The decay time of the oscillations however is approximately $20$ ns. This is much shorter than the 
relaxation time of both the flux-qubit ($T_{1,\rm qb}\sim150$ ns) and the NV$^{-}$ ensemble 
($T_{1,\rm NV} \gg 10 \mu$s).  As we tune the flux-qubit away from the 2.878 GHz resonance, 
the decay time associated with vacuum Rabi measurement becomes significantly longer (Figure 4b). 
From these results, one must conclude that a source of strong dephasing 
of unknown origin exists in the system near resonance. There are several 
likely sources. The most probable is a large electron spin $\frac{1}{2}$ 
bath from the P1 (nitrogen atom substituting a carbon atom) centers 
present in our HPHT Ib-type diamond crystal used to prepare the NV$^{-}$ 
ensemble. In our situation, where there is no external magnetic field, 
the NV$^{-}$ centers and P1 centers naturally couple\cite{Marcos10}. 
Hanson et. al \cite{Hanson08} have shown an enhanced decay may result. The P1 center 
issue can be eliminated to a large extent by applying an external 
magnetic field to split the $\left|\pm 1 \right \rangle$ NV$^{-}$ 
states. A 1 mT field could split these by approximately 15 MHz, 
detuning the P1 centers and thus significantly improving the dephasing 
time of the coupled system. We can also decrease the number of P1 
centers in the sample (from 100 ppm to 1 ppm) by using different synthesized diamond crystals. 
In addition, by using non HPHT Ib-type crystals we can remove the effect of other natural defects that may be 
present. Finally there is also a strong hyperfine interaction ($\sim 100$ MHz) between the  NV$^{-}$ electron spin and $^{13}$C nuclear 
spins. Without the nuclear spins being initially polarized, unwanted 
dephasing will result. By polarizing the nuclear spins this source of 
dephasing can be removed. This should allow us to observe vacuum Rabi 
oscillations where we are limited by $T_2$ of the flux-qubit.

In conclusion, we have experimentally demonstrated  strong coherent coupling 
between a flux-qubit and an ensemble of nitrogen-vacancy color 
centers in single crystal diamond.  Furthermore, we have observed, via vacuum 
Rabi oscillations, the coherent exchange (transfer) of a single quantum of energy. 
This is the first step towards the realization of a long lived quantum memory for 
condensed matter systems with an additional potential future application as an 
interface between the microwave and optical domains.

\noindent {\em Acknowledgments}:  We thank T. Tawara, H. Gotoh and T. Sogawa for optical measurement in the early stage of this work. 
This work was supported in part by the Funding Program for World-Leading Innovative R{\&}D on Science and Technology (FIRST), 
Scientific Research of Specially Promoted Research, Grant No.18001002 by MEXT, 
Grant-in-Aid for Scientific Research on Innovative Areas Grant No.22102502,  
and Scientific Research(A) Grant No.22241025 by Japanese Society for the Promotion of Science (JSPS). 
M.S.E. was supported with a JSPS fellowship.

\end{document}